\documentclass[reprint,superscriptaddress,showpacs,amsmath,pre,longbibliography]{revtex4-1}
\usepackage{graphicx}
\usepackage{subfigure}
\usepackage{txfonts}
\usepackage[colorlinks,
linkcolor=blue,citecolor=blue,anchorcolor=blue,urlcolor=blue]{hyperref}

\begin{document}

\title{Percolation on networks with weak and heterogeneous dependency}

\author{Ling-Wei Kong}
\affiliation{School of the Gifted Young, University of Science and Technology of China, Hefei, 230026, P. R. China}
\author{Ming Li}
\email{minglichn@ustc.edu.cn}
\affiliation{School of Engineering Science, University of Science and Technology of China, Hefei, 230026, P. R. China}
\author{Run-Ran Liu}
\email{runranliu@163.com}
\affiliation{Alibaba Research Center for Complexity Sciences, Hangzhou Normal University, Hangzhou, 311121, P. R. China}
\author{Bing-Hong Wang}
\affiliation{Department of Modern Physics, University of Science and Technology of China, Hefei, 230026, P. R. China}

\date{\today}

\begin{abstract}
In real networks, the dependency between nodes is ubiquitous; however, the dependency is not always complete and homogeneous. In this paper, we propose a percolation model with weak and heterogeneous dependency; i.e., dependency strengths could be different between different nodes. We find that the heterogeneous dependency strength will make the system more robust, and for various distributions of dependency strengths both continuous and discontinuous percolation transitions can be found. For Erd\H{o}s-R\'{e}nyi networks, we prove that the crossing point of the continuous and discontinuous percolation transitions is dependent on the first five moments of the dependency strength distribution. This indicates that the discontinuous percolation transition on networks with dependency is determined not only by the dependency strength but also by its distribution. Furthermore, in the area of the continuous percolation transition, we also find that the critical point depends on the first and second moments of the dependency strength distribution. To validate the theoretical analysis, cases with two different dependency strengths and Gaussian distribution of dependency strengths are presented as examples.

\end{abstract}

\maketitle

\section{Introduction}

The percolation model is an important and widely used model in the study of complex networks\cite{Newman2010Networks}. It describes the properties of the clusters formed by some links and nodes. For instance, bond (link) percolation has been used to study the spreading of epidemic disease\cite{PhysRevE.66.016128}, and site (node) percolation has been used to study the structure and the robustness of complex networks\cite{Cohen2010Complex}. On a tree like network, all these models can be solved exactly by the so-called generating function technique\cite{Wilf2006}. In this way, the percolation model can also provide a theoretical approach to study the problems of complex networks.

In recent years, many modified percolation models have been proposed in the study of complex networks, such as $k$-core percolation, $l$-hop percolation, and clique percolation\cite{PhysRevLett.96.040601,PhysRevLett.94.160202,PhysRevE.84.031113,PhysRevE.92.042116}. In these percolation models, different dependency relationships between adjacent nodes have been proposed to represent the correlations between them in the spreading dynamics on complex networks or the organization of complex networks. Beyond the connection, the dependency relationship between nodes without a direct connection has also been taken into account as a part of the complicated interdependency in real systems\cite{Parshani2011,Buldyrev2010}. In general, this type of dependency is presented by the so-called dependency link. Two nodes connected by such a link will fail together if either of them fails (remove from the network). This mechanism makes the failures easier to spread in the network, and the network with dependency links, is much more fragile than those without them\cite{Boccaletti20141}. More interesting, different from the continuous transition found in the classic percolation, this percolation process demonstrates a discontinuous phase transition. Therefore, this new model has attracted the attention of many physicists\cite{Boccaletti20141}. It is found that there is actually a hybrid phase transition in these models, meaning that the order parameter has a jump at the transition point but there are also critical phenomena related to it\cite{PhysRevLett.109.248701,PhysRevE.90.012803,PhysRevE.93.042109}.

Those researches have shown that the dependency between nodes within a network or between different networks can greatly jeopardize the stability of the whole system. However, most networks in the real world do not seem to be so fragile. Thus, many previous researches are concerning with networks with only a fraction of nodes having dependency links, the so-called partial dependency\cite{PhysRevLett.105.048701,PhysRevE.85.016112,PhysRevE.87.052812}. They found that a reduction in the number of dependency links would make the system more robust, and a crossover of the continuous and discontinuous percolation transitions can be found.

Here we focus on another aspect of this problem. We think that one of the key reasons for the robustness of the real dependent networks is that the strength of dependency is limited and heterogeneous. The failure of a node's dependency partner usually can only reduce its function partially, instead of destroying it completely. For example, in a financial network, when a company loses its partner with a funding requirement, it often only loses some trading links instead of going into bankruptcy. In this paper we study this type of dependency. In our model, a failed node will not cause the node depending on it to fail completely but only to lose some of its connectivity links. Based on this mechanism, both homogeneous and heterogeneous dependencies are considered in this paper.

The paper is organized as follows. In Sec. \ref{model}, the details of our model are given. We will give the results of the general formalism (homogeneous dependency) in Sec. \ref{result1}, and then the heterogeneous dependency is studied in Sec. \ref{result2}. In the last section, we report our conclusions.

\section{Model} \label{model}

Our model displays a network with degree distribution $p_k$ and average degree $\langle k\rangle$. The dependency partners are assigned randomly, and each node has only one dependency partner. As in classical percolation, we occupy a fraction $p$ of the nodes in the network, and all the unoccupied nodes (fraction $1-p$) are considered failed nodes and are removed from the network. When a node loses its dependency partner, each of its links is removed from the network with a probability $\beta$, respectively. Note that $\beta$ could be different for different nodes. The preserved links are no longer affected by the failed dependency partner. This process is called the dependency failure. Besides, all the nodes disconnecting from the giant component are also considered failed nodes and are removed from the network. This process is called connection failure.

Obviously, the dependency failure could start a new connection failure, and vice versa. Therefore, the initial node removal will trigger an alternative occurrence of the dependency and connection failures, called cascading failures. When there is no node or a link removal can be done, the cascading failure ends. The surviving nodes, if any, are called the giant component of the final network. To evaluate the robustness of such systems, we check the fraction of nodes in this giant component, $S$, which marks the size of the giant component for this percolation model. For a given $p$, the larger the giant component $S$ is, the more robust the network is. It is obvious that $S$ can also represent the probability that a randomly chosen node in the original network belongs to the giant component of the final network.

Obviously, the parameter $\beta$ will play a key role in determining the robustness of such networks. For simplification, we call $\beta$ dependency strength. When $\beta=0$ for all nodes, there is no dependency between nodes in the system, and the system just takes a classical percolation on the network. While $\beta=1$ for all nodes, this model reduces to the one discussed by Parshani \emph{et al.}\cite{Parshani2011}, in which the dependency is very strong and the system is fragile. Furthermore, by assigning different nodes with different $\beta$, we can study a system with heterogeneous dependency. This is one of the key issues studied in this paper.

\section{Homogeneous Dependency} \label{result1}

\subsection{General formalism}

In this section, we consider the case of homogeneous dependency; i.e., all nodes have the same dependency strength $\beta$. As mentioned in the last section, the initial node removal will trigger a cascading failure in the network. As a straightforward approach\cite{Buldyrev2010,Parshani2011}, we can calculate the fraction of the preserved nodes for each stage of the cascading failure step by step. In this paper, we use a simpler approach, which considers the final state after the cascades, directly\cite{PhysRevLett.109.248701,Son2012,Li2013}. For this approach, we need to know what kind of nodes can be preserved in the cascading failures. There are two situations: nodes with a working dependency partner, and nodes with a failed dependency partner. For the former, at least one of the node's links must connect to a node in the giant component of the final network. For the latter, some of the node's links will be removed; thus, in the remaining links, at least one link must connect to the giant component of the final network. All we need to do now is to calculate the happening probability of the two cases. It is worth noting that this theoretical method works in the thermodynamic limit, for which the uncertainty caused by the probability $\beta$ in the evolution process does not affect the results.

For convenience, let $R$ be the probability that a link connects to the giant component of the final network. Then, a node with degree $k$ will belong to the giant component with probability $p[1-(1-R)^k]$, where $p$ means that the node cannot be removed in the initial failures. Averaging this probability over the degree distribution of the network, we can get the probability that a randomly chosen node connects to the giant component $p[1-G_0(1-R)]$. Here, $G_0(x)=\sum_{k}p_kx^k$ is the generating function of the degree distribution. Since the dependency partners are paired randomly, such a probability for a randomly chosen node's dependency partner can also be expressed as $p[1-G_0(1-R)]$. Therefore, the probability for the first case can be written as $p^2[1-G_0(1-R)]^2$. For the second case, only the preserved links (with probability $1-\beta$) can be considered; thus, the corresponding probability for survived nodes is $p[1-G_0(1-R+R\beta)]$. Therefore, the size of the giant component $S$ can be written as
\begin{eqnarray}
S &=& p^2[1-G_0(1-R)]^2 \nonumber \\
  & & + p\left[1-G_0(1-R+R\beta)\right]\left[1-p+p G_0(1-R)\right]. \label{s}
\end{eqnarray}
The first term of this equation is for the nodes with a working dependency partner, and the second term is for the nodes with a failed dependency partner.

To solve eq.(\ref{s}), we must get the equation for $R$ first. For this, the generating function $G_{1}(x)=\sum_{k}p_kkx^{k-1}/\langle k\rangle$ is used, which describes the excess degree of the node reached by following a link. In this way, the probability that a link connects to the giant component can be expressed as $p[1-G_1(1-R)]$. This means that at least one of the other links of the node reached by following the link must connect to the giant component. If the link we follow belongs to a node with a failed dependency partner, the probability must be rewritten as $p(1-\beta)[1-G_1(1-R+R\beta)]$. Here, $1-\beta$ means that the link we follow must be preserved when the dependency partner fails. Thus, similar to eq.(\ref{s}), we can get a self-consistent equation for $R$:
\begin{eqnarray}
R &=& p^2[1-G_1(1-R)][1-G_0(1-R)]+ p(1-\beta) \nonumber \\
  & &\times [1-G_1(1-R+R\beta)][1-p+p G_0(1-R)]. \label{r}
\end{eqnarray}
The meanings of the two terms are the same as that in eq.(\ref{s}), but for the node reached by following a link.

\begin{figure}
\centering
\includegraphics[width=0.46\textwidth]{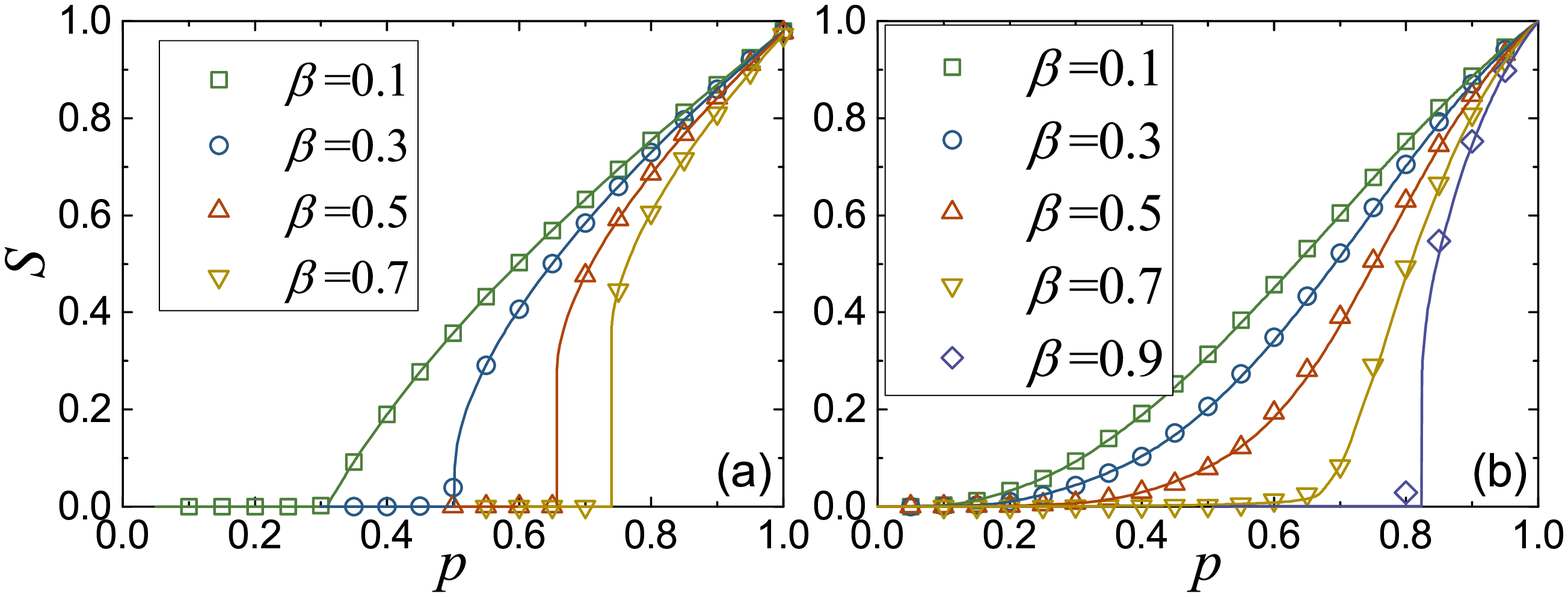}
\caption{The fraction of nodes in the giant component at the end of the cascade process, $S$, is shown as a function of $p$ for different $\beta$. In the simulation, the size of the network is $N=500 000$ for ER networks and $N=50 000$ for SF networks, and the average degree is $\langle k\rangle = 4$. For SF networks, the degree distribution satisfies $p_k\sim k^{-\gamma}$ with $\gamma=2.63$. The corresponding lines are the theoretical results obtained by eqs.(\ref{s}) and (\ref{r}). (a) ER networks, (b) SF networks.}
\label{f1}
\end{figure}

Thus we can obtain $S$ and $R$ from eqs. (\ref{s}) and (\ref{r}) for a network with given $p_k$ and $\beta$. In Fig.\ref{f1}, we give the simulation results for both Erd\H{o}s-R\'{e}nyi (ER) and scale-free (SF) networks, which agree with our theory well. We can also find that the critical point $p_c$ increases with the increasing of dependency strength $\beta$. This indicates that the networks become more robust as $\beta$ decreases. Moreover, Fig.\ref{f1} also indicates that there exists a critical point $\beta_c$, above which the system will show a discontinuous percolation transition.

\subsection{The critical point} \label{cp}

Next, we show how to obtain the critical point $p_c$ and $\beta_c$ of the system. First, we consider the ER network, which takes the degree distribution $p_k=e^{-\langle k\rangle}\langle k\rangle^{k}/k!$. Then, the generating functions $G_0(x)$ and $G_1(x)$ take the same form,
\begin{equation}
G_0(x)=G_1(x)=e^{-\langle k\rangle (1-x)}.
\end{equation}
In this case, eqs. (\ref{s}) and (\ref{r}) can be written as
\begin{eqnarray}
S &=& p^2\left(1-e^{-R\langle k\rangle}\right)^2 \nonumber \\
  & & + p\left[1-e^{-(1-\beta)R\langle k\rangle}\right]\left(1-p+pe^{-R\langle k\rangle}\right), \label{ers} \\
R &=& p^2\left(1-e^{-R\langle k\rangle}\right)^2 \nonumber \\
  & & + p(1-\beta)\left[1-e^{-(1-\beta)R\langle k\rangle}\right]\left(1-p+pe^{-R\langle k\rangle}\right). \label{err}
\end{eqnarray}
From eq.(\ref{ers}), it is easy to know that only a nonzero $R$ can give a meaningful $S$. Thus, we can discusses the solution of eq.(\ref{err}) to obtain the critical point of the system.

\begin{figure}
\centering
\includegraphics[width=0.46\textwidth]{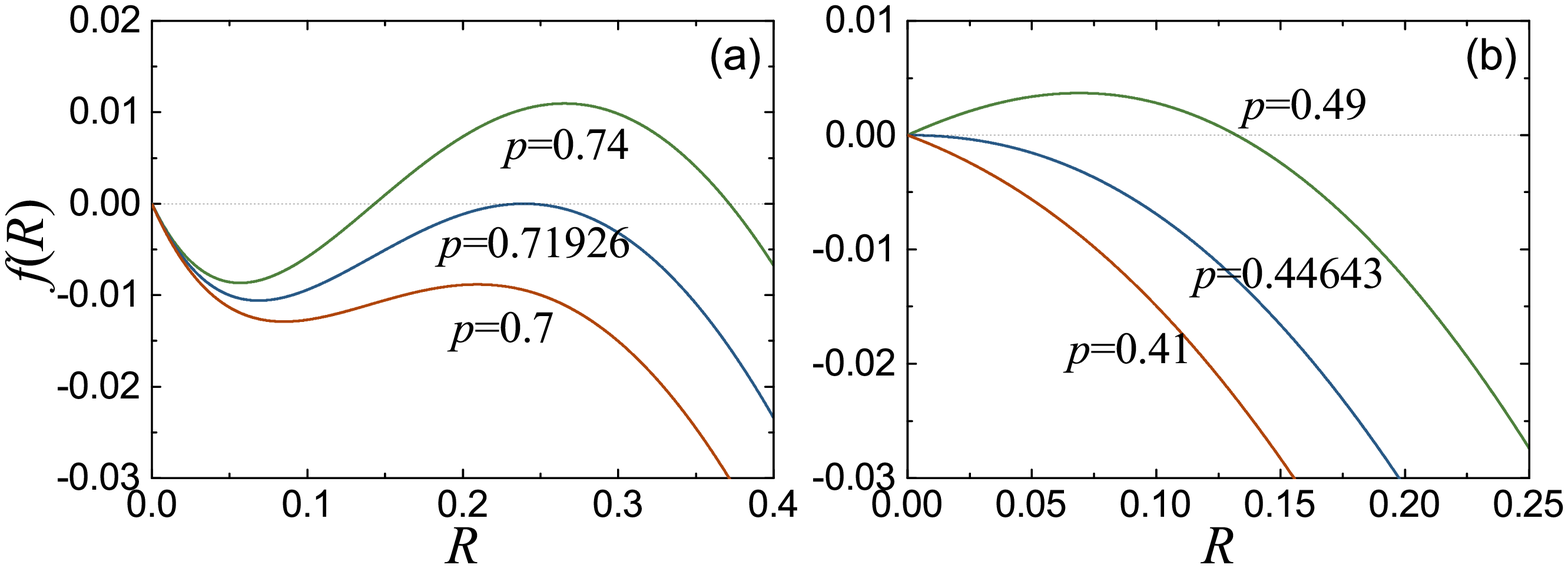}
\caption{Graphical solutions for eq.(\ref{err}) with average degree $\langle k\rangle = 3.5$. (a) $\beta = 0.5$. The percolation transition is discontinuous with the critical point $p_c = 0.71926$ and a nonzero order parameter $R_c>0$. (b) $\beta = 0.2$. The percolation transition is continuous with the critical point $p_c = 0.44643$ and the order parameter $R_c=0$.}
\label{f2}
\end{figure}

Equation (\ref{err}) has a trivial solution at $R=0$, which means that the network is completely fragmented. The nontrivial solution of $R$ can be presented by the crossing points of the curve $f(R)$ defined by eq.(\ref{err}) ($f(R)=rhs-R$, where $rhs$ indicates right-hand side) and $R$-axis as shown in Fig.\ref{f2}. We find that the system shows two different types of solutions with the increasing of $\beta$, corresponding to the two types of percolation transitions shown in Fig.\ref{f1}. For both cases, the critical points of the system satisfy $\partial f(R)/\partial R=0$, which gives
\begin{eqnarray}
&& p_c^2\langle k\rangle e^{-R_c\langle k\rangle}\left[1+\beta -2e^{-R_c\langle k\rangle} +(1-\beta) e^{-(1-\beta)R_c\langle k\rangle}\right] \nonumber \\
&&+p_c\langle k\rangle (1-\beta)^2 e^{-(1-\beta)R_c\langle k\rangle} \left(1-p_c+p_ce^{-R_c\langle k\rangle}\right) =1. \label{fr1}
\end{eqnarray}
For the continuous phase transition, $R_c=0$, the first term of the left-hand side of eq.(\ref{fr1}) vanishes. Thus we can obtain the critical point $p_c^{II}$ of the continuous phase transition,
\begin{equation}
p^{II}_c=\frac{1}{\langle k\rangle (1-\beta)^2}, ~~\beta < \beta_c. \label{pc2}
\end{equation}
When the percolation transition is discontinuous ($\beta >\beta_c$), both terms on the left-hand side of eq.(\ref{fr1}) contribute to the critical point $p_c^{I}$. In this way, we cannot get a closed form of the critical point. However, together with eq.(\ref{err}), we can easily obtain the numerical solution of $p^{I}_c$.

At the critical point $\beta_{c}$, the conditions for the continuous and discontinuous percolation transitions are met, simultaneously. According to the graphical solution shown in Fig \ref{f2}, eq. (\ref{fr1}) gives one solution for the first order region, but two solutions for the second order region. So, at the critical point $\beta_{c}$, the two solutions must merge into one, which corresponds to $(\partial^2f(R)/\partial R^2)|_{R=0}=0$. Together with $R_c=0$, this yields
\begin{equation}
\frac{(1-\beta_c)^3}{p_c}+2(1-\beta_c)^2-2=0.
\end{equation}
At the critical point $\beta_c$, $p_c$ also satisfies eq.(\ref{pc2}), so we have
\begin{equation}
\langle k\rangle(1-\beta_c)^5+2(1-\beta_c)^2-2=0. \label{bc}
\end{equation}
This equation can also be solved numerically. A network with $\beta<\beta_c$ undergoes a continuous percolation transition; otherwise it would be the discontinuous percolation transition.

\begin{figure}
\scalebox{0.3}[0.3]{\includegraphics{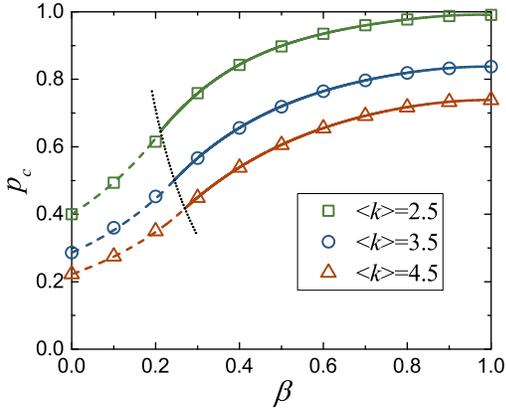}}
\caption{The critical point $p_c$ for different $\beta$ in ER networks. The simulation results are shown by symbols, and the theoretical results obtained by eqs.(\ref{err})-(\ref{pc2}) are shown by solid and dashed lines for the discontinuous and continuous percolation transitions, respectively. The dotted line in the middle separates the discontinuous and continuous percolation transition regions obtained by eq.(\ref{bc}). In the simulation, the size of the network is $N=500 000$.}
\label{f3}
\end{figure}

In Fig.\ref{f3}, we show the phase diagram of the system for different average degrees. As the theory predicts, the phenomenon of crossover in the percolation transition can be found, and the simulation results agree with our theory very well. As it is clear that more links will make the network robust, we can also find that the critical point $p_c$ decreases with the increasing of the average degree $\langle k\rangle$ in Fig.\ref{f3}. In addition, we can also obtain the critical point in a similar way for SF networks; however, we cannot find as simple a form as eqs.(\ref{pc2}) and (\ref{bc}).

\section{Heterogeneous Dependency} \label{result2}

\subsection{Theory}

In reality, the strengths of the dependency between different nodes could be different. In ref.\cite{Li2014}, we have also discussed this problem from the perspective of asymmetric dependency. However, the distribution of the dependency strengths cannot be addressed in such a model. In this model, we can investigate this problem by assigning different nodes with different dependency strengths $\beta$. For the convenience of mathematical treatment, we assume all the dependency strengths of the nodes are chosen from a random variable $\beta=\{\beta_1, \beta_2, \beta_3,\ldots\}$, which follows a distribution $p_{\beta}$. Therefore, we can rewrite eqs.(\ref{s}) and (\ref{r}) as
\begin{eqnarray}
S_\beta &=& p^2[1-G_0(1-R)]^2 \nonumber \\
  & & + p[1-G_0(1-R+R\beta)][1-p+p G_0(1-R)], \label{ns} \\
R &=& p^2[1-G_1(1-R)][1-G_0(1-R)] \nonumber \\
  & & + p[1-p+p G_0(1-R)] \nonumber \\
  && \times \sum_\beta p_\beta(1-\beta)[1-G_1(1-R+R\beta)]. \label{nr}
\end{eqnarray}
Here, $S_\beta$ is the probability that a node with dependency strength $\beta$ belongs to the giant component of the final network. Thus the order parameter of the system is given by $S=\sum_\beta p_\beta S_\beta$.

\begin{figure*}
\centering
\includegraphics[width=0.9\textwidth]{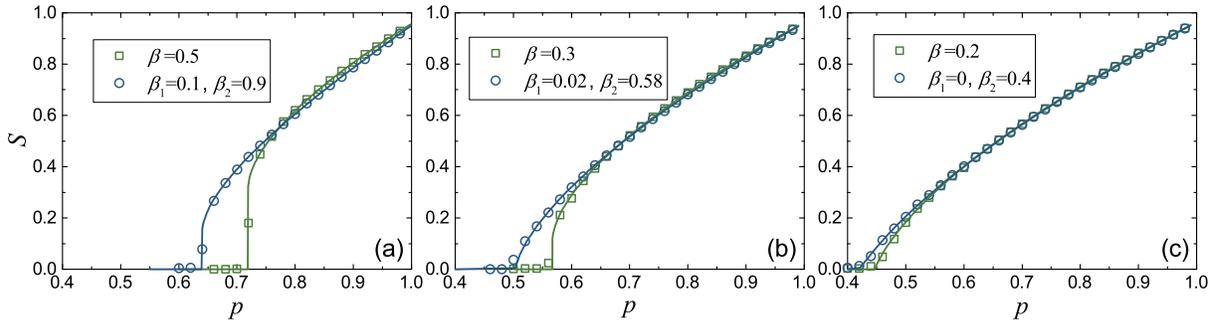}
\caption{The simulation results for ER networks with heterogeneous dependency strength (circles) and homogeneous dependency strength (squares). In the simulation, the fraction $q=0.5$, the size of networks is $N=100 000$ and the average degree $\langle k\rangle=3.5$. The two cases have the same average dependency strength, that is, $q\beta_1+(1-q)\beta_2=\beta$. The solid lines are obtained by eqs.(\ref{ns}) and (\ref{nr}). (a) Both networks undergo the discontinuous percolation transition. (b) The network with heterogeneous dependency strength presents a continuous percolation transition while the one with homogeneous dependency presents a discontinuous percolation transition. (c) Both networks present a continuous percolation transition.}
\label{diff}
\end{figure*}

Using the same method mentioned in the last section, we can obtain the critical point of the continuous percolation transition, $p^{II}_c$, analytically:
\begin{equation}
p^{II}_{c,\beta}=\frac{1}{G_1^\prime(1) \sum_{\beta}p_\beta(1-\beta)^2}=\frac{1}{G_1^\prime(1) \left\langle (1-\beta)^2\right\rangle}. \label{npc2}
\end{equation}
Here, $\langle \cdot \rangle$ means the average over the distribution of the dependency strength $p_\beta$. This equation indicates that the continuous transition point only depends on the first and second moments of the distribution $p_\beta$ for a given network. For homogeneous dependency, the distribution $p_\beta$ can be expressed by a Dirac $\delta$ function $p_\beta=\delta(\beta-\beta_0)$. In this way, eq.(\ref{npc2}) reduces to eq.(\ref{pc2}) for ER networks.

To compare with the case of homogeneous dependency, after substituting $\overline{\beta}=\sum_\beta p_\beta \beta$ into eq.(\ref{pc2}) then we find
\begin{equation}
p^{II}_{c,\overline{\beta}}=\frac{1}{G_1^\prime(1) (1-\sum_\beta p_\beta\beta)^2}=\frac{1}{G_1^\prime(1) \left(1-\langle \beta \rangle\right)^2}.
\end{equation}
It is easy to find that $p^{II}_{c,\beta}$ and $p^{II}_{c,\overline{\beta}}$ satisfy
\begin{equation}
\frac{1}{G_1^\prime(1)}\left(\frac{1}{p^{II}_{c,\beta}}- \frac{1}{p^{II}_{c,\overline{\beta}}}\right) = \langle \beta^{2}\rangle -\langle \beta\rangle^{2}.
\end{equation}
The right-hand side of this equation is just the variance of random variable $\beta$, which always takes a positive value. So for a network with nonzero $G_1^\prime(1)$, we have
\begin{equation}
p^{II}_{c,\beta} < p^{II}_{c,\overline{\beta}}. \label{p2cb}
\end{equation}
This indicates that for the same average dependency strength $\overline{\beta}$, the heterogeneous dependency strength $\beta$ will make the system more robust when the percolation transition is continuous.

For the discontinuous percolation transition, although we cannot get a closed form of the critical points $p^{I}_{c,\beta}$ and $p^{I}_{c,\overline{\beta}}$, for ER networks we can also prove that (see Appendix \ref{app} for details)
\begin{equation}
p^{I}_{c,\beta} < p^{I}_{c,\overline{\beta}}. \label{p1cb}
\end{equation}

In addition, for some $\beta$, the network with heterogeneous dependency strength presents a continuous transition, while the corresponding homogeneous one presents a discontinuous transition (see Fig.\ref{diff}). For this case, the network with heterogeneous dependency strength must also give a smaller critical point. If not, the curves $p_{c,\beta}$ and $p_{c,\overline{\beta}}$ must have two crossing points, at which the type of the percolation transition changes.
This leads to $p^{I}_{c,\beta} = p^{I}_{c,\overline{\beta}}$ or $p^{II}_{c,\beta} = p^{II}_{c,\overline{\beta}}$, obviously contradicting eqs.(\ref{p2cb}) and (\ref{p1cb}). In this way, for a given ER network we always have
\begin{equation}
p_{c,\beta} < p_{c,\overline{\beta}}. \label{pcgt}
\end{equation}
This indicates that for a given average dependency strength, an ER network with heterogeneous dependency strength will always be more robust than a network with homogeneous dependency, whether the percolation transition is continuous or discontinuous.

For heterogeneous dependency, there is no one special dependency strength at which the type of the percolation transition changes as that in the homogeneous case. However, by the similar method used in Sec.\ref{cp}, we can also obtain an equation which meets the conditions of the discontinuous and continuous percolation simultaneously. For ER networks, that is
\begin{equation}
\langle k\rangle\left\langle (1-\beta)^5\right\rangle+2\left\langle (1-\beta)^2 \right\rangle-2=0. \label{bdc}
\end{equation}
This indicates that the crossing point of the discontinuous and continuous percolation transitions is dependent on the first five moments of the distribution of $\beta$. In addition, it is easy to find that for homogeneous dependency this equation will reduce to eq.(\ref{bc}).

As the boundary of the discontinuous and continuous percolation transitions, the left-hand side of eq.(\ref{bdc}) can also be used to determine the type of the percolation transition for a given distribution $p_{\beta}$. We relabel the left-hand side of eq.(\ref{bdc}) as a function of the dependency strength distribution $W(p_{\beta})$. Obviously, $W(p_{\beta})$ is a decreasing function of the moments of distribution $p_\beta$. As the meaning of the dependency strength, a larger moment of distribution $p_\beta$ clearly leads to a poor robustness. Therefore, if a distribution $p_\beta$ gives a $W(p_{\beta})<0$, the corresponding percolation transition is discontinuous; otherwise it is continuous.

\subsection{Examples}

\begin{figure}
\scalebox{0.3}[0.3]{\includegraphics{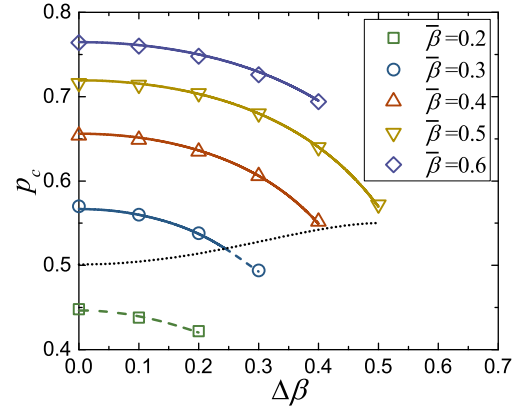}}
\caption{The critical point $p_c$ for ER networks with two dependency strengths. The fraction $q=0.5$; the dependency strengths are $\beta_{1}=\overline{\beta}-\Delta\beta$ and $\beta_{2}=\overline{\beta}+\Delta\beta$. In the simulation, the size of the network is $N=50,000$, and the average degree $\langle k \rangle=3.5$. The solid lines denote the discontinuous percolation transition points while the dashed lines denote that of the continuous percolation transition obtained by eqs.(\ref{nr}) and (\ref{npc2}). The dotted line is the boundary between the two types of transitions obtained by eq.(\ref{bdc}).}
\label{diffzt}
\end{figure}

Next, we give a simple example for the network with heterogeneous dependency strength. We consider the case for which there exist two dependency strengths $\beta_1=\overline{\beta}-\Delta\beta$ (fraction $q$) and $\beta_2=\overline{\beta}+\Delta\beta$ (fraction $1-q$) in an ER network. For this case, $\Delta\beta$ represents the difference of the two dependency strengths $\beta_1$ and $\beta_2$, and the average dependency strength $\overline{\beta}$ is independent of the fraction $q$ and the difference $\Delta\beta$.

As shown in Figs.\ref{diff} and \ref{diffzt}, with the increasing of the difference $\Delta\beta$ of $\beta_1$ and $\beta_2$, the networks become more fragile. This is consistent with our theory; i.e., for the same average dependency strength, the network with heterogeneous dependency strength is more robust than the network with homogeneous dependency strength. From Fig.\ref{diffzt}, we can also find that with increasing $\overline{\beta}$, the robustness of the networks will decrease, and the percolation transition changes from the continuous one to the discontinuous. For a moderate $\overline{\beta}$, both types of percolation transitions can be found in the system for different $\Delta \beta$.

\begin{figure}
\scalebox{0.3}[0.3]{\includegraphics{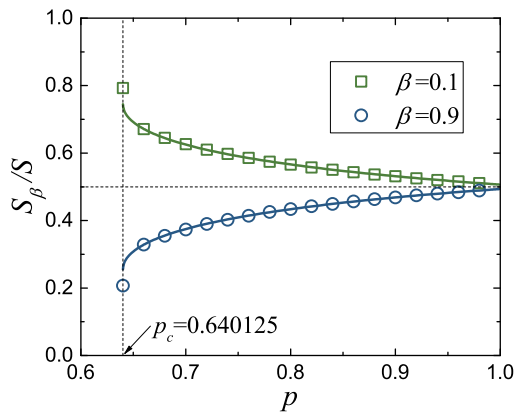}}
\caption{The fractions of nodes with different dependency strengths in the giant component, $S_\beta/S$, as a function of the fraction of the initial preserved nodes $p$. In the simulation, there are two dependency strengths $\beta=0.1$ and $=0.9$ in the system, and mixed uniformly ($q=0.5$). The networks used here are ER networks with size $N=100 000$ and average degree $\langle k\rangle=3.5$. The solid line denotes the theoretical results obtained by eqs.(\ref{ns}) and (\ref{nr}).}
\label{diff2}
\end{figure}

From our theory, we can also obtain the fraction of each kind of nodes that survive in the final state, $S_\beta/S$. From Fig.\ref{diff2}, we can find that the nodes with the smaller dependency strengths more easily survive the cascading failures, and hit the extreme at the critical point $p_c$. With the increase of $p$, this fraction converges to the initial fraction of the node with dependency strength $\beta$. This could provide us a method to identify the critical point for the case of heterogeneous dependency in the simulation.

\begin{figure}
\scalebox{0.3}[0.3]{\includegraphics{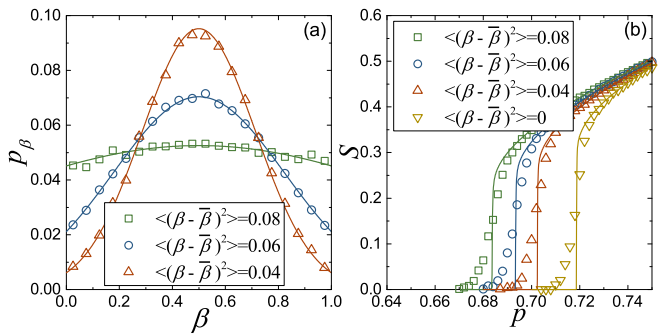}}
\caption{The simulation results for ER networks with Gaussian distributions of dependency strengths. In the simulation, the size of networks is $N=50 000$ and the average degree $\langle k\rangle=3.5$. The average dependency strength is $\overline{\beta}=0.5$, $\sigma$ and $\lambda$ are set as $0.21355, 0.3225, 0.9061$ and $0.9508, 0.879, 0.4189$, respectively, corresponding to $\langle(\beta-\overline{\beta})^2\rangle=0.04, 0.06, 0.08$. (a) The frequency counting of the dependency strengths generated by $p(\beta)$ with bin size $\delta\beta=0.05$. The solid lines are obtained by $p_\beta=p(\beta)\delta\beta$.  (b) The size of the giant component $S$ as a function of the fraction of the initial preserved nodes $p$ for different second moments of the distribution of dependency strengths. The solid lines are obtained by eqs.(\ref{ns}), (\ref{nr}), and (\ref{pb}).}
\label{fp}
\end{figure}

For a more general distribution of dependency strengths, we consider a case for which the dependency strengths $\beta$ are chosen from a random variable, whose probability density distribution is Gaussian:
\begin{equation}
p(\beta)=\frac{1}{\lambda} \frac{1}{\sqrt{2\pi}\sigma}e^{-\frac{(\beta-\overline{\beta})^2}{2\sigma^2}}. \label{pb}
\end{equation}
Here, $\overline{\beta}$ is the average and $\sigma^2$ is the variance. Since dependency strength $\beta$ ranges from zero to $1$, a factor $\lambda$ is used to renormalize this distribution. By choosing different $\sigma$ and $\lambda$, we can obtain a set of dependency strengths with a given second moment $\langle(\beta-\overline{\beta})^2\rangle$. Note that $p(\beta)$ is a probability density distribution and $p_\beta$ is the probability distribution. To use $p(\beta)$, eq. ($\ref{nr}$) must be written in integral form. For the discrete form, $p_\beta=p(\beta)\delta\beta$, where $\delta\beta$ is the bin size used in frequency counting.

In Fig.\ref{fp}, we show the simulation results for different second moments of the distribution of dependency strengths $\langle(\beta-\overline{\beta})^2\rangle$. We can find that all these results are in agreement with the theoretical prediction. As we know, the second moment of a random variable indicates the heterogeneity of its distribution. The larger the second moment is, the more heterogeneous the distribution is. Therefore, as the second moment $\langle(\beta-\overline{\beta})^2\rangle$ increases, the network becomes more robust. This theoretical result is also confirmed by the simulation results shown in Fig.\ref{fp}.

\section{Conclusions}

In this paper we have proposed a model to study the percolation process on networks with limited dependency between nodes. As the usual setting of networks with dependency, the failure of one node will affect the function of its dependency node. However, instead of destroying the node completely, in our model the failed node only leads to the failures of its dependency partner's links. Specifically, when a node fails, each link of its dependency partner will fail with a probability $\beta$, respectively.

By assigning different nodes with different dependency strengths $\beta$, we generally compare the systems with homogeneous and heterogeneous dependency strengths. We find that for a given average dependency strength, the heterogeneous dependency strength will make the system more robust than the homogeneous dependency strength, and both the continuous and discontinuous percolation transitions can be found for different dependency strength distributions. This indicates that the type of the percolation transition on networks with dependency is not only determined by the dependency strength but also its distribution. Furthermore, for ER networks we prove that the crossing point of the continuous and discontinuous percolation transitions is dependent on the first five moments of the distribution of dependency strengths. All these findings indicate that the distribution of dependency strengths plays an important role in the robustness of networks, and more research is needed for a deeper understanding of the heterogeneous dependency.

\begin{acknowledgments}
The research of M.L. was supported by the National Natural Science Foundation of China under Grant No.61503355. The research of R.-R.L. was supported by the National Natural Science Foundation of China under Grant No.11305042. The research of B.-H.W. was supported by the National Natural Science Foundation of China under Grant No.11275186.
\end{acknowledgments}

\appendix

\section{Proof for eq.(\ref{p1cb})} \label{app}
Similar to the discussion shown in Fig.\ref{f2}, we define
\begin{eqnarray}
f(R,\beta,p) &=& p^2(1-\xi)^2 + p(1-\beta)(1-p+p\xi)[1-\zeta(\beta)] \nonumber \\
&&-R, \label{f}
\end{eqnarray}
where $\xi=e^{-R\langle k\rangle}$ and $\zeta(\beta)=e^{-(1-\beta)R\langle k\rangle}$. It is clear that both $\xi$ and $\zeta(\beta)$ are smaller than $1$. Then, for a network with a distribution $p_\beta$, such a function will be the linear combination of eq.(\ref{f}) with different $\beta$, i.e.,
\begin{eqnarray}
F(R,\beta,p) &=& \sum_{\beta}p_\beta f(R,\beta,p) \nonumber \\
&=& p(1-p+p\xi)\sum_\beta p_\beta(1-\beta)[1-\zeta(\beta)] \nonumber \\
&&+p^2(1-\xi)^2 -R. \label{ff}
\end{eqnarray}
Similar to that shown in Fig.\ref{f2}, the tangency points of eqs.(\ref{f}) and (\ref{ff}) with the $R$-axis correspond to the critical points of the two systems, respectively.

\begin{figure}
\scalebox{0.3}[0.3]{\includegraphics{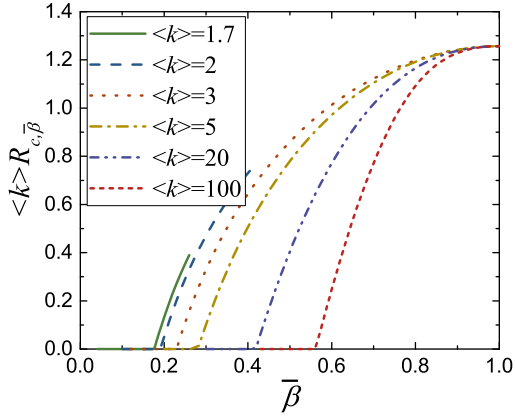}}
\caption{The parameter $\langle k\rangle R_{c,\overline{\beta}}$ plotted as a function of the dependency strength $\overline{\beta}$ for different average degrees. The results are obtained by eqs.(\ref{err}) and (\ref{fr1}). The discontinuous percolation corresponds to the area in which $R_{c,\overline{\beta}}>0$.}
\label{rcb}
\end{figure}

If the heterogeneous case gives a smaller critical point, $F(R,\beta,p)$ must be larger than $f(R,\overline{\beta},p)$ when $p$ is close to the critical point. Here, $\overline{\beta}=\sum_\beta p_\beta \beta$. That is,
\begin{eqnarray}
&& F(R,\beta,p)-f(R,\overline{\beta},p)  \nonumber \\
&=& p(1-p+p\xi)\left[ (1-\overline{\beta})\zeta(\overline{\beta}) -\sum_\beta p_\beta(1-\beta)\zeta(\beta)\right]  \nonumber \\
&>& 0. \label{gt}
\end{eqnarray}
It is easy to know that if
\begin{eqnarray}
g(\beta)&=&(1-\beta)\zeta(\beta) \nonumber \\
&=&(1-\beta)e^{-\langle k\rangle(1-\beta)R}
\end{eqnarray}
is a concave function, eq. (\ref{gt}) will be satisfied for all $\beta$. Next, we focus on this.

The first and second derivatives of $g(\beta)$ are
\begin{eqnarray}
\frac{dg(\beta)}{d\beta} &=& \zeta(\beta)\left[\langle k\rangle R (1-\beta)-1\right], \\
\frac{d^2g(\beta)}{d\beta^2} &=& \langle k\rangle R \zeta(\beta)\left[\langle k\rangle R(1-\beta)-2\right].
\end{eqnarray}
For a concave function, $d^2g(\beta)/d\beta^2> 0$ for all $\beta$, that is $\langle k\rangle R> 2$. Actually, eq. (\ref{gt}) only needs to be met at the critical point $p_{c,\overline{\beta}}$ given by $f(R,\overline{\beta},p)$\footnote{In factor, any $R$ satisfying $F(R,\beta,p_{c,\overline{\beta}})>0$ is enough to demonstrate that $F(R,\beta,p)$ gives a smaller critical point than $f(R,\overline{\beta},p)$. The method and the choice $R=R_{c,\overline{\beta}}$ used here is just one of the ways to find such an $R$.}. So we simply need to prove $R_{c,\overline{\beta}}< 2/\langle k\rangle$.

For $\overline{\beta}=1$, the function $f(R,\overline{\beta},p)$ gives the critical point $R_{c,\overline{\beta}} \approx 1.26/\langle k\rangle$, which satisfies the condition discussed above. Unfortunately, we cannot get a closed form of $R_{c,\overline{\beta}}$ for $\overline{\beta}<1$. However, from the numerical solution shown in Fig.\ref{rcb}, we can find that $R_{c,\overline{\beta}}$ is a monotonous increasing function of $\overline{\beta}$ in the area of the discontinuous percolation. As the theory predicts, $\langle k\rangle R_{c,\overline{\beta}}$ for different average degrees all converge to $1.26$ at $\overline{\beta}=1$. Thus, we address $R_{c,\overline{\beta}}< 2/\langle k\rangle$; i.e., $g(\beta)$ is a concave function at the critical point $p_{c,\overline{\beta}}$.

Above all, we have proved that $F(R_{c,\overline{\beta}},\beta,p_{c,\overline{\beta}})-f(R_{c,\overline{\beta}},\overline{\beta},p_{c,\overline{\beta}})$ is always positive. Since $f(R_{c,\overline{\beta}},\overline{\beta},p_{c,\overline{\beta}})=0$, we obtain $F(R_{c,\overline{\beta}},\beta,p_{c,\overline{\beta}})>0$. In addition, we know that both $f(R,\overline{\beta},p)$ and $F(R,\beta,p)$ are continuous functions. Therefore, we conclude that with the increasing of $p$, before the curve $f(R,\overline{\beta},p)$ touches the $R$-axis, the curve $F(R,\beta,p)$ has already had some crossing points with the $R$-axis. This is just the meaning of eq.(\ref{p1cb}).

\bibliography{ref}

\begin{thebibliography}{21}%
\makeatletter
\providecommand \@ifxundefined [1]{%
 \@ifx{#1\undefined}
}%
\providecommand \@ifnum [1]{%
 \ifnum #1\expandafter \@firstoftwo
 \else \expandafter \@secondoftwo
 \fi
}%
\providecommand \@ifx [1]{%
 \ifx #1\expandafter \@firstoftwo
 \else \expandafter \@secondoftwo
 \fi
}%
\providecommand \natexlab [1]{#1}%
\providecommand \enquote  [1]{``#1''}%
\providecommand \bibnamefont  [1]{#1}%
\providecommand \bibfnamefont [1]{#1}%
\providecommand \citenamefont [1]{#1}%
\providecommand \href@noop [0]{\@secondoftwo}%
\providecommand \href [0]{\begingroup \@sanitize@url \@href}%
\providecommand \@href[1]{\@@startlink{#1}\@@href}%
\providecommand \@@href[1]{\endgroup#1\@@endlink}%
\providecommand \@sanitize@url [0]{\catcode `\\12\catcode `\$12\catcode
  `\&12\catcode `\#12\catcode `\^12\catcode `\_12\catcode `\%12\relax}%
\providecommand \@@startlink[1]{}%
\providecommand \@@endlink[0]{}%
\providecommand \url  [0]{\begingroup\@sanitize@url \@url }%
\providecommand \@url [1]{\endgroup\@href {#1}{\urlprefix }}%
\providecommand \urlprefix  [0]{URL }%
\providecommand \Eprint [0]{\href }%
\providecommand \doibase [0]{http://dx.doi.org/}%
\providecommand \selectlanguage [0]{\@gobble}%
\providecommand \bibinfo  [0]{\@secondoftwo}%
\providecommand \bibfield  [0]{\@secondoftwo}%
\providecommand \translation [1]{[#1]}%
\providecommand \BibitemOpen [0]{}%
\providecommand \bibitemStop [0]{}%
\providecommand \bibitemNoStop [0]{.\EOS\space}%
\providecommand \EOS [0]{\spacefactor3000\relax}%
\providecommand \BibitemShut  [1]{\csname bibitem#1\endcsname}%
\let\auto@bib@innerbib\@empty
\bibitem [{\citenamefont {Newman}(2010)}]{Newman2010Networks}%
  \BibitemOpen
  \bibfield  {author} {\bibinfo {author} {\bibfnamefont {Mark E.~J.}\
  \bibnamefont {Newman}},\ }\href@noop {} {\emph {\bibinfo {title} {Networks:
  An Introduction}}}\ (\bibinfo  {publisher} {Oxford University Press},\
  \bibinfo {year} {2010})\BibitemShut {NoStop}%
\bibitem [{\citenamefont {Newman}(2002)}]{PhysRevE.66.016128}%
  \BibitemOpen
  \bibfield  {author} {\bibinfo {author} {\bibfnamefont {M.~E.~J.}\
  \bibnamefont {Newman}},\ }\bibfield  {title} {\enquote {\bibinfo {title}
  {Spread of epidemic disease on networks},}\ }\href {\doibase
  10.1103/PhysRevE.66.016128} {\bibfield  {journal} {\bibinfo  {journal} {Phys.
  Rev. E}\ }\textbf {\bibinfo {volume} {66}},\ \bibinfo {pages} {016128}
  (\bibinfo {year} {2002})}\BibitemShut {NoStop}%
\bibitem [{\citenamefont {Cohen}\ and\ \citenamefont
  {Havlin}(2010)}]{Cohen2010Complex}%
  \BibitemOpen
  \bibfield  {author} {\bibinfo {author} {\bibfnamefont {Reuven}\ \bibnamefont
  {Cohen}}\ and\ \bibinfo {author} {\bibfnamefont {Shlomo}\ \bibnamefont
  {Havlin}},\ }\href@noop {} {\emph {\bibinfo {title} {Complex Networks:
  Structure, Robustness and Function}}}\ (\bibinfo  {publisher} {Cambridge
  University Press},\ \bibinfo {year} {2010})\BibitemShut {NoStop}%
\bibitem [{\citenamefont {Wilf}(2006)}]{Wilf2006}%
  \BibitemOpen
  \bibfield  {author} {\bibinfo {author} {\bibfnamefont {Herbert~S.}\
  \bibnamefont {Wilf}},\ }\href@noop {} {\emph {\bibinfo {title}
  {Generatingfunctionology}}}\ (\bibinfo  {publisher} {Taylor \& Francis
  Ltd.},\ \bibinfo {year} {2006})\BibitemShut {NoStop}%
\bibitem [{\citenamefont {Dorogovtsev}\ \emph {et~al.}(2006)\citenamefont
  {Dorogovtsev}, \citenamefont {Goltsev},\ and\ \citenamefont
  {Mendes}}]{PhysRevLett.96.040601}%
  \BibitemOpen
  \bibfield  {author} {\bibinfo {author} {\bibfnamefont {S.~N.}\ \bibnamefont
  {Dorogovtsev}}, \bibinfo {author} {\bibfnamefont {A.~V.}\ \bibnamefont
  {Goltsev}}, \ and\ \bibinfo {author} {\bibfnamefont {J.~F.~F.}\ \bibnamefont
  {Mendes}},\ }\bibfield  {title} {\enquote {\bibinfo {title} {$k$-core
  organization of complex networks},}\ }\href {\doibase
  10.1103/PhysRevLett.96.040601} {\bibfield  {journal} {\bibinfo  {journal}
  {Phys. Rev. Lett.}\ }\textbf {\bibinfo {volume} {96}},\ \bibinfo {pages}
  {040601} (\bibinfo {year} {2006})}\BibitemShut {NoStop}%
\bibitem [{\citenamefont {Der\'enyi}\ \emph {et~al.}(2005)\citenamefont
  {Der\'enyi}, \citenamefont {Palla},\ and\ \citenamefont
  {Vicsek}}]{PhysRevLett.94.160202}%
  \BibitemOpen
  \bibfield  {author} {\bibinfo {author} {\bibfnamefont {Imre}\ \bibnamefont
  {Der\'enyi}}, \bibinfo {author} {\bibfnamefont {Gergely}\ \bibnamefont
  {Palla}}, \ and\ \bibinfo {author} {\bibfnamefont {Tam\'as}\ \bibnamefont
  {Vicsek}},\ }\bibfield  {title} {\enquote {\bibinfo {title} {Clique
  percolation in random networks},}\ }\href {\doibase
  10.1103/PhysRevLett.94.160202} {\bibfield  {journal} {\bibinfo  {journal}
  {Phys. Rev. Lett.}\ }\textbf {\bibinfo {volume} {94}},\ \bibinfo {pages}
  {160202} (\bibinfo {year} {2005})}\BibitemShut {NoStop}%
\bibitem [{\citenamefont {Shang}\ \emph {et~al.}(2011)\citenamefont {Shang},
  \citenamefont {Luo},\ and\ \citenamefont {Xu}}]{PhysRevE.84.031113}%
  \BibitemOpen
  \bibfield  {author} {\bibinfo {author} {\bibfnamefont {Yilun}\ \bibnamefont
  {Shang}}, \bibinfo {author} {\bibfnamefont {Weiliang}\ \bibnamefont {Luo}}, \
  and\ \bibinfo {author} {\bibfnamefont {Shouhuai}\ \bibnamefont {Xu}},\
  }\bibfield  {title} {\enquote {\bibinfo {title} {$l$-hop percolation on
  networks with arbitrary degree distributions and its applications},}\ }\href
  {\doibase 10.1103/PhysRevE.84.031113} {\bibfield  {journal} {\bibinfo
  {journal} {Phys. Rev. E}\ }\textbf {\bibinfo {volume} {84}},\ \bibinfo
  {pages} {031113} (\bibinfo {year} {2011})}\BibitemShut {NoStop}%
\bibitem [{\citenamefont {Li}\ \emph {et~al.}(2015)\citenamefont {Li},
  \citenamefont {Deng},\ and\ \citenamefont {Wang}}]{PhysRevE.92.042116}%
  \BibitemOpen
  \bibfield  {author} {\bibinfo {author} {\bibfnamefont {Ming}\ \bibnamefont
  {Li}}, \bibinfo {author} {\bibfnamefont {Youjin}\ \bibnamefont {Deng}}, \
  and\ \bibinfo {author} {\bibfnamefont {Bing-Hong}\ \bibnamefont {Wang}},\
  }\bibfield  {title} {\enquote {\bibinfo {title} {Clique percolation in random
  graphs},}\ }\href {\doibase 10.1103/PhysRevE.92.042116} {\bibfield  {journal}
  {\bibinfo  {journal} {Phys. Rev. E}\ }\textbf {\bibinfo {volume} {92}},\
  \bibinfo {pages} {042116} (\bibinfo {year} {2015})}\BibitemShut {NoStop}%
\bibitem [{\citenamefont {Parshani}\ \emph {et~al.}(2011)\citenamefont
  {Parshani}, \citenamefont {Buldyrev},\ and\ \citenamefont
  {Havlin}}]{Parshani2011}%
  \BibitemOpen
  \bibfield  {author} {\bibinfo {author} {\bibfnamefont {Roni}\ \bibnamefont
  {Parshani}}, \bibinfo {author} {\bibfnamefont {Sergey~V}\ \bibnamefont
  {Buldyrev}}, \ and\ \bibinfo {author} {\bibfnamefont {Shlomo}\ \bibnamefont
  {Havlin}},\ }\bibfield  {title} {\enquote {\bibinfo {title} {Critical effect
  of dependency groups on the function of networks},}\ }\href {\doibase
  10.1073/pnas.1008404108} {\bibfield  {journal} {\bibinfo  {journal} {Proc.
  Natl. Acad. Sci. U.S.A.}\ }\textbf {\bibinfo {volume} {108}},\ \bibinfo
  {pages} {1007—1010} (\bibinfo {year} {2011})}\BibitemShut {NoStop}%
\bibitem [{\citenamefont {Buldyrev}\ \emph {et~al.}(2010)\citenamefont
  {Buldyrev}, \citenamefont {Parshani}, \citenamefont {Paul}, \citenamefont
  {Stanley},\ and\ \citenamefont {Havlin}}]{Buldyrev2010}%
  \BibitemOpen
  \bibfield  {author} {\bibinfo {author} {\bibfnamefont {Sergey~V.}\
  \bibnamefont {Buldyrev}}, \bibinfo {author} {\bibfnamefont {Roni}\
  \bibnamefont {Parshani}}, \bibinfo {author} {\bibfnamefont {Gerald}\
  \bibnamefont {Paul}}, \bibinfo {author} {\bibfnamefont {H.~Eugene}\
  \bibnamefont {Stanley}}, \ and\ \bibinfo {author} {\bibfnamefont {Shlomo}\
  \bibnamefont {Havlin}},\ }\bibfield  {title} {\enquote {\bibinfo {title}
  {Catastrophic cascade of failures in interdependent networks},}\ }\href
  {http://dx.doi.org/10.1038/nature08932} {\bibfield  {journal} {\bibinfo
  {journal} {Nature}\ }\textbf {\bibinfo {volume} {464}},\ \bibinfo {pages}
  {1025--1028} (\bibinfo {year} {2010})}\BibitemShut {NoStop}%
\bibitem [{\citenamefont {Boccaletti}\ \emph {et~al.}(2014)\citenamefont
  {Boccaletti}, \citenamefont {Bianconi}, \citenamefont {Criado}, \citenamefont
  {del Genio}, \citenamefont {G\'omez-Garde{\~n}es}, \citenamefont {Romance},
  \citenamefont {Sendi{\~n}a-Nadal}, \citenamefont {Wang},\ and\ \citenamefont
  {Zanin}}]{Boccaletti20141}%
  \BibitemOpen
  \bibfield  {author} {\bibinfo {author} {\bibfnamefont {S.}~\bibnamefont
  {Boccaletti}}, \bibinfo {author} {\bibfnamefont {G.}~\bibnamefont
  {Bianconi}}, \bibinfo {author} {\bibfnamefont {R.}~\bibnamefont {Criado}},
  \bibinfo {author} {\bibfnamefont {C.I.}\ \bibnamefont {del Genio}}, \bibinfo
  {author} {\bibfnamefont {J.}~\bibnamefont {G\'omez-Garde{\~n}es}}, \bibinfo
  {author} {\bibfnamefont {M.}~\bibnamefont {Romance}}, \bibinfo {author}
  {\bibfnamefont {I.}~\bibnamefont {Sendi{\~n}a-Nadal}}, \bibinfo {author}
  {\bibfnamefont {Z.}~\bibnamefont {Wang}}, \ and\ \bibinfo {author}
  {\bibfnamefont {M.}~\bibnamefont {Zanin}},\ }\bibfield  {title} {\enquote
  {\bibinfo {title} {The structure and dynamics of multilayer networks},}\
  }\href {\doibase 10.1016/j.physrep.2014.07.001} {\bibfield  {journal}
  {\bibinfo  {journal} {Phys. Rep.}\ }\textbf {\bibinfo {volume} {544}},\
  \bibinfo {pages} {1--122} (\bibinfo {year} {2014})},\ \bibinfo {note} {the
  structure and dynamics of multilayer networks}\BibitemShut {NoStop}%
\bibitem [{\citenamefont {Baxter}\ \emph {et~al.}(2012)\citenamefont {Baxter},
  \citenamefont {Dorogovtsev}, \citenamefont {Goltsev},\ and\ \citenamefont
  {Mendes}}]{PhysRevLett.109.248701}%
  \BibitemOpen
  \bibfield  {author} {\bibinfo {author} {\bibfnamefont {G.~J.}\ \bibnamefont
  {Baxter}}, \bibinfo {author} {\bibfnamefont {S.~N.}\ \bibnamefont
  {Dorogovtsev}}, \bibinfo {author} {\bibfnamefont {A.~V.}\ \bibnamefont
  {Goltsev}}, \ and\ \bibinfo {author} {\bibfnamefont {J.~F.~F.}\ \bibnamefont
  {Mendes}},\ }\bibfield  {title} {\enquote {\bibinfo {title} {Avalanche
  collapse of interdependent networks},}\ }\href {\doibase
  10.1103/PhysRevLett.109.248701} {\bibfield  {journal} {\bibinfo  {journal}
  {Phys. Rev. Lett.}\ }\textbf {\bibinfo {volume} {109}},\ \bibinfo {pages}
  {248701} (\bibinfo {year} {2012})}\BibitemShut {NoStop}%
\bibitem [{\citenamefont {Zhou}\ \emph {et~al.}(2014)\citenamefont {Zhou},
  \citenamefont {Bashan}, \citenamefont {Cohen}, \citenamefont {Berezin},
  \citenamefont {Shnerb},\ and\ \citenamefont {Havlin}}]{PhysRevE.90.012803}%
  \BibitemOpen
  \bibfield  {author} {\bibinfo {author} {\bibfnamefont {Dong}\ \bibnamefont
  {Zhou}}, \bibinfo {author} {\bibfnamefont {Amir}\ \bibnamefont {Bashan}},
  \bibinfo {author} {\bibfnamefont {Reuven}\ \bibnamefont {Cohen}}, \bibinfo
  {author} {\bibfnamefont {Yehiel}\ \bibnamefont {Berezin}}, \bibinfo {author}
  {\bibfnamefont {Nadav}\ \bibnamefont {Shnerb}}, \ and\ \bibinfo {author}
  {\bibfnamefont {Shlomo}\ \bibnamefont {Havlin}},\ }\bibfield  {title}
  {\enquote {\bibinfo {title} {Simultaneous first- and second-order percolation
  transitions in interdependent networks},}\ }\href {\doibase
  10.1103/PhysRevE.90.012803} {\bibfield  {journal} {\bibinfo  {journal} {Phys.
  Rev. E}\ }\textbf {\bibinfo {volume} {90}},\ \bibinfo {pages} {012803}
  (\bibinfo {year} {2014})}\BibitemShut {NoStop}%
\bibitem [{\citenamefont {Lee}\ \emph {et~al.}(2016)\citenamefont {Lee},
  \citenamefont {Choi}, \citenamefont {Stippinger}, \citenamefont {Kert\'esz},\
  and\ \citenamefont {Kahng}}]{PhysRevE.93.042109}%
  \BibitemOpen
  \bibfield  {author} {\bibinfo {author} {\bibfnamefont {Deokjae}\ \bibnamefont
  {Lee}}, \bibinfo {author} {\bibfnamefont {S.}~\bibnamefont {Choi}}, \bibinfo
  {author} {\bibfnamefont {M.}~\bibnamefont {Stippinger}}, \bibinfo {author}
  {\bibfnamefont {J.}~\bibnamefont {Kert\'esz}}, \ and\ \bibinfo {author}
  {\bibfnamefont {B.}~\bibnamefont {Kahng}},\ }\bibfield  {title} {\enquote
  {\bibinfo {title} {Hybrid phase transition into an absorbing state:
  Percolation and avalanches},}\ }\href {\doibase 10.1103/PhysRevE.93.042109}
  {\bibfield  {journal} {\bibinfo  {journal} {Phys. Rev. E}\ }\textbf {\bibinfo
  {volume} {93}},\ \bibinfo {pages} {042109} (\bibinfo {year}
  {2016})}\BibitemShut {NoStop}%
\bibitem [{\citenamefont {Parshani}\ \emph {et~al.}(2010)\citenamefont
  {Parshani}, \citenamefont {Buldyrev},\ and\ \citenamefont
  {Havlin}}]{PhysRevLett.105.048701}%
  \BibitemOpen
  \bibfield  {author} {\bibinfo {author} {\bibfnamefont {Roni}\ \bibnamefont
  {Parshani}}, \bibinfo {author} {\bibfnamefont {Sergey~V.}\ \bibnamefont
  {Buldyrev}}, \ and\ \bibinfo {author} {\bibfnamefont {Shlomo}\ \bibnamefont
  {Havlin}},\ }\bibfield  {title} {\enquote {\bibinfo {title} {Interdependent
  networks: Reducing the coupling strength leads to a change from a first to
  second order percolation transition},}\ }\href {\doibase
  10.1103/PhysRevLett.105.048701} {\bibfield  {journal} {\bibinfo  {journal}
  {Phys. Rev. Lett.}\ }\textbf {\bibinfo {volume} {105}},\ \bibinfo {pages}
  {048701} (\bibinfo {year} {2010})}\BibitemShut {NoStop}%
\bibitem [{\citenamefont {Dong}\ \emph {et~al.}(2012)\citenamefont {Dong},
  \citenamefont {Gao}, \citenamefont {Tian}, \citenamefont {Du},\ and\
  \citenamefont {He}}]{PhysRevE.85.016112}%
  \BibitemOpen
  \bibfield  {author} {\bibinfo {author} {\bibfnamefont {Gaogao}\ \bibnamefont
  {Dong}}, \bibinfo {author} {\bibfnamefont {Jianxi}\ \bibnamefont {Gao}},
  \bibinfo {author} {\bibfnamefont {Lixin}\ \bibnamefont {Tian}}, \bibinfo
  {author} {\bibfnamefont {Ruijin}\ \bibnamefont {Du}}, \ and\ \bibinfo
  {author} {\bibfnamefont {Yinghuan}\ \bibnamefont {He}},\ }\bibfield  {title}
  {\enquote {\bibinfo {title} {Percolation of partially interdependent networks
  under targeted attack},}\ }\href {\doibase 10.1103/PhysRevE.85.016112}
  {\bibfield  {journal} {\bibinfo  {journal} {Phys. Rev. E}\ }\textbf {\bibinfo
  {volume} {85}},\ \bibinfo {pages} {016112} (\bibinfo {year}
  {2012})}\BibitemShut {NoStop}%
\bibitem [{\citenamefont {Zhou}\ \emph {et~al.}(2013)\citenamefont {Zhou},
  \citenamefont {Gao}, \citenamefont {Stanley},\ and\ \citenamefont
  {Havlin}}]{PhysRevE.87.052812}%
  \BibitemOpen
  \bibfield  {author} {\bibinfo {author} {\bibfnamefont {Di}~\bibnamefont
  {Zhou}}, \bibinfo {author} {\bibfnamefont {Jianxi}\ \bibnamefont {Gao}},
  \bibinfo {author} {\bibfnamefont {H.~E.}\ \bibnamefont {Stanley}}, \ and\
  \bibinfo {author} {\bibfnamefont {Shlomo}\ \bibnamefont {Havlin}},\
  }\bibfield  {title} {\enquote {\bibinfo {title} {Percolation of partially
  interdependent scale-free networks},}\ }\href {\doibase
  10.1103/PhysRevE.87.052812} {\bibfield  {journal} {\bibinfo  {journal} {Phys.
  Rev. E}\ }\textbf {\bibinfo {volume} {87}},\ \bibinfo {pages} {052812}
  (\bibinfo {year} {2013})}\BibitemShut {NoStop}%
\bibitem [{\citenamefont {Son}\ \emph {et~al.}(2012)\citenamefont {Son},
  \citenamefont {Bizhani}, \citenamefont {Christensen}, \citenamefont
  {Grassberger},\ and\ \citenamefont {Paczuski}}]{Son2012}%
  \BibitemOpen
  \bibfield  {author} {\bibinfo {author} {\bibfnamefont {Seung-Woo}\
  \bibnamefont {Son}}, \bibinfo {author} {\bibfnamefont {Golnoosh}\
  \bibnamefont {Bizhani}}, \bibinfo {author} {\bibfnamefont {Claire}\
  \bibnamefont {Christensen}}, \bibinfo {author} {\bibfnamefont {Peter}\
  \bibnamefont {Grassberger}}, \ and\ \bibinfo {author} {\bibfnamefont {Maya}\
  \bibnamefont {Paczuski}},\ }\bibfield  {title} {\enquote {\bibinfo {title}
  {Percolation theory on interdependent networks based on epidemic
  spreading},}\ }\href {http://stacks.iop.org/0295-5075/97/i=1/a=16006}
  {\bibfield  {journal} {\bibinfo  {journal} {Europhys. Lett.}\ }\textbf
  {\bibinfo {volume} {97}},\ \bibinfo {pages} {16006} (\bibinfo {year}
  {2012})}\BibitemShut {NoStop}%
\bibitem [{\citenamefont {Li}\ \emph {et~al.}(2013)\citenamefont {Li},
  \citenamefont {Liu}, \citenamefont {Jia},\ and\ \citenamefont
  {Wang}}]{Li2013}%
  \BibitemOpen
  \bibfield  {author} {\bibinfo {author} {\bibfnamefont {Ming}\ \bibnamefont
  {Li}}, \bibinfo {author} {\bibfnamefont {Run-Ran}\ \bibnamefont {Liu}},
  \bibinfo {author} {\bibfnamefont {Chun-Xiao}\ \bibnamefont {Jia}}, \ and\
  \bibinfo {author} {\bibfnamefont {Bing-Hong}\ \bibnamefont {Wang}},\
  }\bibfield  {title} {\enquote {\bibinfo {title} {Critical effects of
  overlapping of connectivity and dependence links on percolation of
  networks},}\ }\href {http://stacks.iop.org/1367-2630/15/i=9/a=093013}
  {\bibfield  {journal} {\bibinfo  {journal} {New J. Phys.}\ }\textbf {\bibinfo
  {volume} {15}},\ \bibinfo {pages} {093013} (\bibinfo {year}
  {2013})}\BibitemShut {NoStop}%
\bibitem [{\citenamefont {Li}\ \emph {et~al.}(2014)\citenamefont {Li},
  \citenamefont {Liu}, \citenamefont {Jia},\ and\ \citenamefont
  {Wang}}]{Li2014}%
  \BibitemOpen
  \bibfield  {author} {\bibinfo {author} {\bibfnamefont {Ming}\ \bibnamefont
  {Li}}, \bibinfo {author} {\bibfnamefont {Run-Ran}\ \bibnamefont {Liu}},
  \bibinfo {author} {\bibfnamefont {Chun-Xiao}\ \bibnamefont {Jia}}, \ and\
  \bibinfo {author} {\bibfnamefont {Bing-Hong}\ \bibnamefont {Wang}},\
  }\bibfield  {title} {\enquote {\bibinfo {title} {Cascading failures on
  networks with asymmetric dependence},}\ }\href
  {http://stacks.iop.org/0295-5075/108/i=5/a=56002} {\bibfield  {journal}
  {\bibinfo  {journal} {Europhys. Lett.}\ }\textbf {\bibinfo {volume} {108}},\
  \bibinfo {pages} {56002} (\bibinfo {year} {2014})}\BibitemShut {NoStop}%
\bibitem [{Note1()}]{Note1}%
  \BibitemOpen
  \bibinfo {note} {In factor, any $R$ satisfying $F(R,\beta ,p_{c,\protect
  \overline {\beta }})>0$ is enough to demonstrate that $F(R,\beta ,p)$ gives a
  smaller critical point than $f(R,\protect \overline {\beta },p)$. The method
  and the choice $R=R_{c,\protect \overline {\beta }}$ used here is just one of
  the ways to find such an $R$.}\BibitemShut {Stop}%
\end{thebibliography}%

\end{document}